# Macroscopic yarns of FeCl$_3$-intercalated collapsed carbon nanotubes with high doping and stability

Cristina Madrona,[1,2] María Vila,[1] Freddy E. Oropeza,[3] Víctor A. de la Peña O'Shea,[3] Juan J. Vilatela[1,*]

1 IMDEA Materials Institute, Eric Kandel 2, 28906 Getafe, Madrid, Spain

2 Facultad de Ciencias, Universidad Autónoma de Madrid, Francisco Tomás y Valiente, 7, 28049 Madrid, Spain

3 IMDEA Energy Institute, Avda. Ramón de la Sagra 3, 28935 Móstoles, Madrid, Spain

* juanjose.vilatela@imdea.org

**ABSTRACT.** Macroscopic arrays of highly-crystalline nanocarbons offer the possibility of modifying the electronic structure of their low dimensional constituents, for example through doping, and studying the resulting collective bulk behaviour. Insertion of electron donors or acceptors between graphitic layers is an attractive method to reversibly increase charge carrier concentration without disruption of the sp2-conjugated system. This work demonstrates FeCl$_3$ intercalation into fibres made up of collapsed (flattened) carbon nanotubes. The bundles of collapsed CNTs, similar to crystallites of graphitic nanoribbons, host elongated layered FeCl$_3$ crystals of hundreds of nm long, much longer than previous reports on graphitic materials and directly observable by transmission electron microscopy and X-ray diffraction. Intercalated CNT fibres remain stable after months of exposure to ambient conditions, partly due to the spontaneous formation of passivating monolayers of FeClO at the crystal edge, preventing both desorption of intercalant and further hydrolysis. Raman spectroscopy shows substantial electron transfer from the CNTs to FeCl$_3$, a well-known acceptor, as observed by G band upshifts as large as 25 cm$^{-1}$. After resolving Raman features for the inner and outer layers of the collapsed CNTs, strain and dynamic effect contributions of charge transfer to the Raman upshift could be decoupled, giving a Fermi level downshift of –0.72 eV and a large average free carrier concentration of 5.3·10$^{13}$ cm$^{-2}$ (0.014 electrons per carbon atom) in the intercalated system. Four-probe resistivity measurements show an increase in conductivity by a factor of six upon intercalation.

## 1. INTRODUCTION

Carbon nanotubes (CNTs) remain a fascinating building block for macroscopic materials. When highly crystalline and of few layers, they can combine unrivalled mechanical properties [1,2], thermal conductivity on par with diamond [3] and charge carrier mobility above silicon [4]. Transferring these properties to bulk materials has progressed significantly in recent years through improved control of the molecular structure at the point of synthesis and of the assembly process into aligned fibres of long CNT bundles. Tensile mechanical properties of CNT fibres are routinely above reference carbon fibres [5], and electrical [6,7] and thermal [6,8] conductivities rival those of copper (on a mass basis).

An attractive feature of macroscopic fibres of CNTs is the possibility to change the electronic structure of the individual building blocks to intervene on the bulk fibre properties, for example through doping or chemical functionalisation. There is particular interest in methods that modify the CNT electronic structure or increase the carrier density while preserving the high charge carriers mobility of the sp2-conjugated system. This can be achieved by adsorption of dopants on the external surface of the bundles [9], by filling CNTs [10], or through intercalation between CNTs in bundles [11]. In intercalation of CNTs, electron acceptors/donors (oxidising/reducing species) are inserted between graphitic planes of neighbouring CNTs through a charge transfer process between the dopant and the CNT host, analogous to graphite intercalation compounds (GICs) [12]. In the intercalated state, single-wall carbon nanotubes (SWCNTs) have shown new optical absorption bands with Br$_2$, Cs and AuCl$_3$ at saturation doping [13,14], a decrease in work function, e.g., with AuCl$_3$ [14], and simultaneous increase in electrical conductivity and mechanical properties in I$_2$ intercalated CNT fibres [7]. GICs additionally show superconductivity, e.g., in KC$_8$ and CaC$_6$ compounds [15], increased energy storage capacities by using FeCl$_3$-intercalated graphites as anode [16–18], among other properties. Indeed, producing continuous, macroscopic fibres of intercalated CNTs with the rich properties of GICs has been an attractive prospect since the early days of CNT research [19].

Methods to produce GICs, transferrable to macroscopic fibres of CNTs, include intercalation in liquid-phase, electrochemical intercalation and gas-phase intercalation. There has been particular focus on concentrated acids (H$_2$SO$_4$), which can protonate ultra-high purity SWCNTs/DWCNTs and form thermodynamic solutions or lyotropic dispersions that can be spun into continuous fibres [20]. Doping by $H_2SO_4$ increases bulk electrical conductivity by a factor of 5-10, to levels above most metals on a mass basis [6], highlighting one motivation for intercalating CNT fibres. X-ray measurements confirm that $H_2SO_4$ can intercalate by insertion in the triangular SWCNTs lattice of the bundles in the fibres [20]. Electron transfer from (to) the CNTs in the intercalated state produces hardening (softening) of the tangential vibrational modes and changes in electron-phonon coupling, which translate into shifts of the Raman G band [21,22]. For $H_2SO_4$/SWCNT, the large upshift of 30 cm$^{-1}$ in the $G^+$ Raman band [23] is indicative of a significant charge transfer. Fibre of more generic CNT types can be intercalated electrochemically with Li$^+$ ions. In the intercalated state, CNT fibre anodes show a 5-fold increase in longitudinal electrical conductivity, although with a smaller G band upshift of around 10 cm$^{-1}$ and with progressive degradation upon electrochemical cycling [24].

Vapour intercalation of SWNTs in powder form was reported with different electron donors (such as K, Cs, or Rb) [21,25,26] and acceptors (such as $Br_2$, $I_2$, $FeCl_3$) [21,27,28] about two decades ago. Such works were constrained to test systems of ultra-high crystallinity SWCNTs with very narrow diameter distribution in triangular lattice arrangement, typically produced by arc discharge or laser ablation. Following this work, macroscopic fibres of CNTs exposed to vapour of $I_2$ [7,29], $Br_2$ [30], and ICl [29] have shown large increases in electrical conductivity through doping, but without evidence of intercalation or the large degree of charge transfer observed in GICs.

Overall, it remains a challenge to intercalate acceptor and donors able to produce large charge transfer throughout the entire macroscopic fibres of conventional CNTs. This requires, firstly, improving our current understanding of the structure of intercalated CNTs in terms of the role of CNT morphology (shape, number of layers), surface chemistry and bundling structure. It is unclear, for example, whether homogeneous intercalation is restricted to samples of CNTs associated in bundles with a crystalline superlattice, or whether a high degree of sp2 conjugation is sufficient. Furthermore, the stability of GICs is known to be affected by structural perfection of the graphitic layers, with deintercalation rates going from minutes [31] to months for some compounds [32,33]; but in CNTs, stacking and longitudinal (in-plane) perfection are in fact decoupled.

In this work, we produce fibres of collapsed double-wall CNTs intercalated with $FeCl_3$, a p-type dopant leading to large charge transfer in GICs. The bundles of collapsed DWCNTs, similar to graphitic nanoribbons, host elongated bi- and three-dimensional layered $FeCl_3$ crystals of hundreds of nanometres in length. The large lateral size of the intercalant inserted between parallel collapsed CNTs enables its stabilisation in ambient conditions, together with the spontaneous formation of a passivating oxide layer of FeClO. Large charge transfer and increased bulk electrical conductivity of the fibres remain even after months exposed to ambient conditions.

## 2. EXPERIMENTAL SECTION

### 2.1 CNT fibre fabrication

Collapsed CNTs were prepared by floating-catalyst chemical vapour deposition (FC-CVD) at 1250 °C, using toluene, ferrocene and thiophene as carbon source, catalyst and promoter, respectively. At an elemental ratio C:Fe:S of 98.7:0.8:0.5, the precursors favour the formation of CNTs of large diameter and few layers, which thus flatten and collapse. The constituent carbon nanotubes are effectively closed and have two layers on average. CNT fibres were used for intercalation without purification. The samples consisted of unidirectional fabrics of CNT fibres wound directly from the reactor.

### 2.2. Method of intercalation and experimental set-up

Intercalation of anhydrous ferric chloride in collapsed carbon nanotube fibres was achieved by the two-zone vapour transport technique. First, sample and intercalant were introduced in a tubular glass capsule separated by a certain distance to allow heating them at different temperatures afterwards. This process is done in argon atmosphere in a glove box, where the system is subsequently sealed at a low degree of vacuum (~10$^{-2}$ mbar). The capsule was then placed in a horizontal tubular furnace such that the difference in temperature between the zones is ~10 °C, with 310 °C in the CNT fibre zone. Exposure to vapour was maintained for two days, without external chlorine supply.

In order to stop the intercalation reaction while avoiding condensation of $FeCl_3$ in the sample surface, the part of the tube closer to the intercalant was first quenched to force condensation of the remaining gas in that cold zone, while the sample zone remained at a temperature higher than the condensation temperature of $FeCl_3$. Inspection of samples by electron microscopy confirmed that this method was successful at avoiding condensation in the exterior surface of the bundles. After the treatment, the samples were extracted from the glass capsule in a glove box and stored in argon for analysis. A subset of samples was extracted from the glove box and kept under ambient conditions for stability studies.

### 2.3. Materials characterization

High resolution transmission electron microscopy (HRTEM) images were taken in a JEOL JEM 2100 operating at 200 kV. XRD measurements correspond to radial profiles obtained from 2D wide-angle x-ray scattering (WAXS) patterns obtained at the NCD-SWEET beamline of ALBA synchrotron. The presented patterns result from azimuthal integration of the WAXS data after background subtraction. Multiple individual patterns with a short exposure of 3 s were collected in order to minimise possible sample disruption. No evidence of beam-induced damage or structural evolution was observed under these conditions after multiple pattern collection. In order to minimise exposure to humid air, one of the samples ("fresh") was tested in an Ar-filled Kapton cell built in house. X-ray photoelectron spectra (XPS) were recorded in a lab-based spectrometer (SPECS GmbH, Berlin) using a monochromatic Al Kα1 source (hv=1486.6 eV) operating at 50 W. X-rays are microfocused onto a spot size on the sample of 300 μm in diameter. The analyser is a SPECS PHOIBOS 150 NAP, a 180° hemispherical energy analyser with 150 mm mean radius. The total energy resolution of the measurements was about 0.5 eV. The binding energy scale was calibrated against the Ag Fermi level. Raman spectroscopy was carried out in a Renishaw inVia micro-Raman spectrometer with a laser wavelength of 532 nm (2.33 eV), and in a Bruker Senterra with 785 nm wavelength (1.58 eV), using 50x objectives and a low power configuration in order to avoid heating effects and sample disruption. Electrical resistance was measured with a 2450 Keithley source–measurement unit using four probe contact electrodes. Samples consisted of two types: CNT films 5 mm x 3.5 μm in section and 4 cm long and individual filaments. To enable a direct comparison of the effect of intercalation on longitudinal conductivity, measurements were performed on the same samples before and after intercalation. Different zones of the sample were tested to prove homogeneity. Sample thickness was measured with a profilometer; no appreciable changes occurred after intercalation.

## 3. RESULTS AND DISCUSSION

The material used for intercalation consists of fabrics of continuous fibres made up of long bundles of collapsed (i.e., flattened) CNTs stacked in graphitic-like domains. **Figure 1** and Figure S1 present examples of transmission electron micrographs showing their bundle structure, composed of mainly 2-4 walls collapsed CNTs stacked turbostratically into graphitic ribbon-like structures. The largest bundles contain around 15-45 flattened nanotubes, forming stacks ~10-30 nm thick. The mean width of the collapsed CNTs is ~15 nm. Collapsed nanotubes can be thought of as CNTs in which two opposite walls of the tube attach, forming a flattened structure. Larger diameter and fewer layers, i.e., a lower bending rigidity, make the collapsed state of CNTs more stable than a round morphology, particularly in bundles, where they also maximise external contact [34–36]. Seen as a graphitic crystal, the collapsed CNT bundles synthesised in this work form very long domains with no grain boundaries (tubes length is hundreds of microns), offering large graphitic areas to host and stabilise intercalants. Overall, with the view of forming intercalation compounds, this system has structural attributes distinct from graphite and carbon fibres. The sp2 carbon layers have large in-plane crystallite size, but with a relatively low interlayer binding energy due to turbostratic stacking. In addition, the elongated crystals offer lateral access to the intercalant.

These fibres of collapsed CNTs were intercalated with $FeCl_3$ using a two-zone vapour transport method [37,38]. Briefly, the CNT fibre was continuously exposed to $FeCl_3$ vapour for two days under anhydrous conditions, while maintaining a higher temperature at the sample than the source to favour intercalation and avoid condensation (see Experimental Section).

### 3.1. Morphological characterization of the intercalation compound

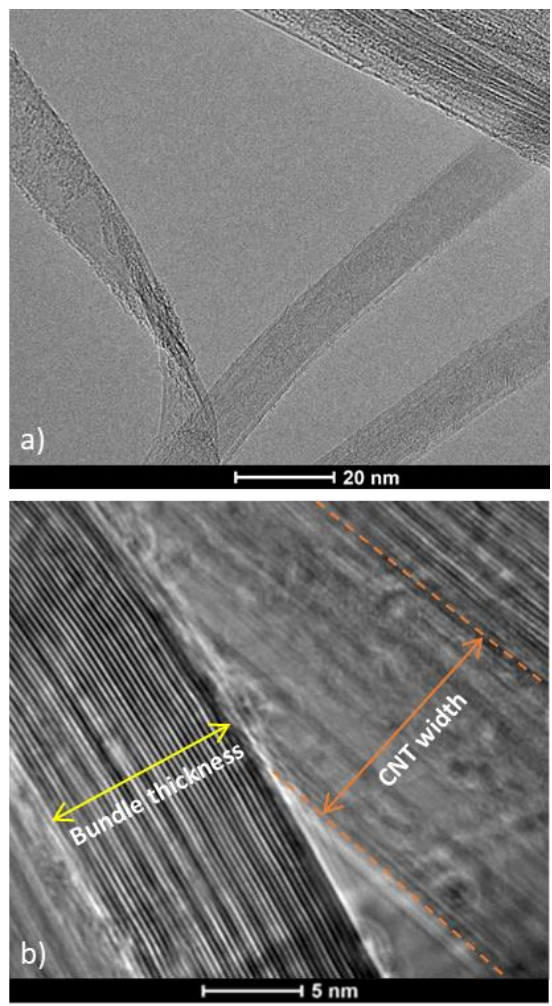

**Figure 1.** a) HRTEM image of a bundle composed of double walled collapsed CNTs. b) HRTEM micrograph showing two bundles of collapsed CNTs forming highly graphitic stacked domains. The width of the collapsed tubes is around 15 nm.

The ribbon-like structure of the collapsed CNT bundles enables direct observation of intercalation domains by high resolution transmission electron microscopy (HRTEM) imaging, which provides extraordinary new insight into the structure of the intercalated system. Several aspects of the morphology of the intercalation compound are shown in **Figure 2**. The extension of the intercalated $FeCl_3$ monolayers is large, as observed in **Figure 2**.a. In fact, these domains are so long as to make it challenging to follow them under TEM and accurately determine their size, but lengths well over 200 nm are found throughout the CNT fibre material. These domains are much longer than previously observed even in highly crystalline traditional carbons such as highly-oriented pyrolytic graphite (80 nm) [39] and benzene-derived carbon fibres heat treated at 2800 °C (40 nm) [32]. Intercalated regions are readily distinguished in TEM micrographs due to the expanded separation between carbon planes. The distance between neighbouring collapsed CNTs when a $FeCl_3$ monolayer is intercalated increases typically from ~3.45 Å to ~9.4 Å (see Figure S2 and inset in **Figure 2**.a), which corresponds closely to the separation observed in $FeCl_3$-GICs [17]. A separation of 11.3 Å is also observed in some regions, which is due to the conversion of FeCl$_3$ into FeClO at some of the edges, as discussed later.

Although intercalated FeCl$_3$ predominates as a monolayer, interestingly, some areas present multilayer crystals of this intercalant. **Figure 2**.b shows examples of domains with 5 layers of FeCl$_3$ intercalated between CNTs. A closer inspection of an intercalation edge (**Figure 2**.b) shows a terrace-like structure, suggesting that multilayer crystals are formed over an already intercalated monolayer, i.e., once the carbon planes had already been separated. In this scenario, entry of a new FeCl$_3$ nucleus in an intercalated region does not require overcoming the van der Waals interaction between carbon planes, but only the interaction between the charged FeCl$_3$ and sp2 carbon layers, in addition to the strain energy from additional CNT deformation. HRTEM observation also shows that multilayers of FeCl$_3$ are crystalline and have crystallographic registry. Ordered (003) planes, highlighted with arrows in **Figure 2**.c, are parallel to the planar walls of the collapsed CNTs. Interference patterns of this (003) line with the (2-13) and (2-1-3) are also observed in a small region, pointing to the corresponding crystal orientation shown as inset in **Figure 2**.c. Interestingly, we have observed that when the intercalant is present as a multilayer, the distance between the outermost FeCl$_3$ layer and its adjacent carbon layer is reduced to ~4.3 Å (Figure S3), lower than the sum of half the interlayer distances of FeCl$_3$ (5.65 Å) and of the graphitic planes (3.45 Å). This decreased distance may result from a high interaction between FeCl$_3$ and carbon layers in the intercalated state.

Based on the observed intercalant morphology and bundle structure, we envisage an intercalation process in similar steps as FeCl$_3$-GICs, but with faster intercalation kinetics. First, FeCl$_3$ clusters adsorb between CNT edges on the lateral face of the bundles. Once a nucleus or island is created, it enters as a monolayer between the collapsed CNTs, overcoming their van der Waals interaction through electron transfer from the carbon material to the FeCl$_3$ acceptor compound. Entry of the FeCl$_3$ nucleus acts as a wedge, forming a separation between CNTs that propagates laterally beyond the initial nucleus size because of the high in-plane stiffness of the CNTs. Bundle wedging is favoured by the relatively small bundle thickness; in graphite, for example, high compaction of graphitic planes in a monolithic structure difficult their separation, forcing the intercalation process to start by the uppermost layers [40]. Furthermore, the inherently turbostratic stack makes the separation forces required for intercalation lower compared to AB stacked graphite. In addition, the fact that bundle edges are curved graphitic layers rather than reactive crystal ends is likely to favour intercalation instead of alternative reaction paths, such as chlorination. Once the wedge is created, the opening between collapsed CNTs is readily accessible for further lateral entry of FeCl$_3$, thus leading to longitudinal growth of the intercalated domain.

In addition to the morphology discussed above, some intercalated domains show the presence of FeClO. They can be identified under TEM observation due to the increase in CNT layer distance to around 11.3 Å (see **Figure 3**.a and inset in **Figure 2**.a), which corresponds to the sum of the interlayer distances in common FeClO crystals (7.95 Å) and in these collapsed CNTs (3.45 Å). Nevertheless, observation of intercalated bundles in HRTEM micrographs through Fast Fourier Transform (FFT) confirms the predominance of FeCl$_3$ in the inner part, through the appearance of the (20) basal reflections from FeCl$_3$ throughout the intercalated bundle and appreciable features from FeClO only at the edges, as shown in **Figure 3**.a. These observations indicate that the small oxide domains at the bundle edge have a passivating function; they prevent the inner FeCl$_3$ layers from deintercalation and from further hydrolysis. A similar passivating role of FeClO in stopping insertion of water molecules has also been observed in FeCl$_3$-GICs, where it is often found at the edges of graphite crystals directly exposed to moist air [40]. In the case of CNT bundles, the main difference is the large availability of crystal edges. The structural evolution during the complete process is schematised in **Figure 3**.b, showing progression from a pristine bundle of collapsed CNTs, to intercalated FeCl$_3$ and to the formation of passivating domains of FeClO.

Further insights into the structure and stability of the intercalated material was obtained from x-ray diffraction (XRD) measurements. **Figure 3**.c presents XRD patterns for a pristine sample, and for intercalated samples kept under two different conditions, one stored and measured in a nominally inert atmosphere ("fresh"), and another stored in ambient conditions for two months. The main peak of interest in the pristine sample is a broad (002) reflection centred at q~1.82 Å$^{-1}$, resulting from an interplanar distance (d=2π/q) typical of turbostratically stacked graphitic layers (~3.45 Å). In the FeCl$_3$-intercalated "fresh" sample (red pattern in **Figure 3**.a), the appearance of crystalline (hkl) reflections of FeCl$_3$ indicates the presence of intercalated multilayers with an ordered structure, in agreement with TEM micrographs (**Figure 2**.b). The most prominent reflection is the (003), which gives a distance between consecutive basal planes of FeCl$_3$ in intercalated multilayers of 5.65 Å, smaller than the one appearing in free FeCl$_3$ crystals (5.8 Å). The peak at 1.46 Å$^{-1}$, corresponding to an interplanar distance of 4.3 Å, is assigned to the reflection between carbon layer and adjacent intercalated FeCl$_3$ when multilayers are introduced. This assignation is supported by selected area electron diffraction (SAED) measurements in oriented bundles, showing that this reflection is aligned with diffraction spots coming from the (002) of the CNTs and the (003) of FeCl$_3$; and also by direct measurement of this distance in HRTEM micrographs, as mentioned above. An example of SAED pattern and of HRTEM image corresponding to this situation are presented in Figure S3. By other side, some of the unassigned peaks in the XRD pattern probably correspond to the different stages present in the sample. By analysing twenty TEM micrographs, 82% of times we observe stage 4 and the rest 18% correspond to stage 6. We also observe random sequences due to the nature of the sample, meaning that not all collapsed tubes are double-walled, and thus we see patterns such as 2 walls-intercalant-4 walls-intercalant-6 walls-intercalant, and so on. Also unintercalated zones are observed, as typical in FeCl$_3$-GICs, no matter what dominant stage appears. Finally, there are also interplanar reflections of FeClO in the XRD pattern of the "fresh" sample; an indication that some exposure to humid air occurred.

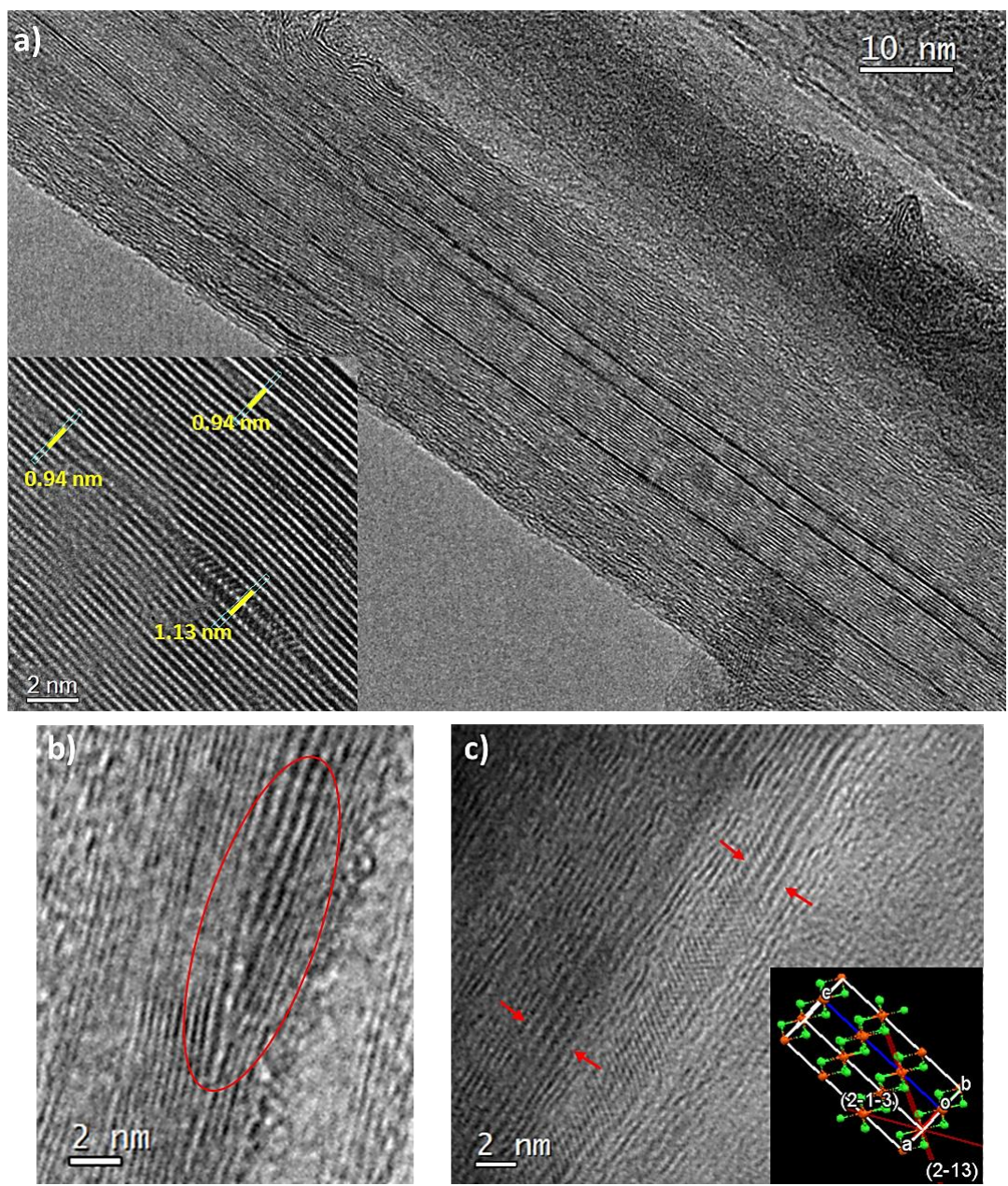

**Figure 2.** HRTEM micrographs of a sample of FeCl$_3$-intercalated collapsed CNT fibre with minimised exposure to air. a) Long intercalated regions are observed as dark stripes at domains with an increased distance between graphitic planes. b) Multilayers of FeCl$_3$ intercalated between collapsed nanotubes. An intercalated monoloayer can act as a wedge that favours the formation of multiple layers of intercalant. c) Multilayers of intercalant show crystallographic registry between layers, observed as interlayer features in the Fast Fourier Transform (FFT) of the interference pattern, and illustrated in the model in the inset.

An intercalated sample exposed to ambient conditions for two months was also analysed by XRD and TEM. The XRD pattern (**Figure 3**.c) contains peaks of FeClO both in the form of intercalated multilayers (interlayer reflections present) and also monolayers. The monolayers give rise to a peak at q of 0.56 Å$^{-1}$ (see inset in **Figure 3**.c), resulting from an interplanar distance of 11.24 Å corresponding to the distance between carbon layers with an intercalated FeClO monolayer, in agreement with TEM. But the most intense peaks come from multilayer FeClO reflections, for example the (101). Peaks from FeCl$_3$ are also observed, but they are weaker and mainly from monolayers. This suggests preferential conversion of multi-layers to FeClO, which may be due to the stronger interaction of FeCl$_3$ with CNT layers as a monolayer favouring its stabilisation.

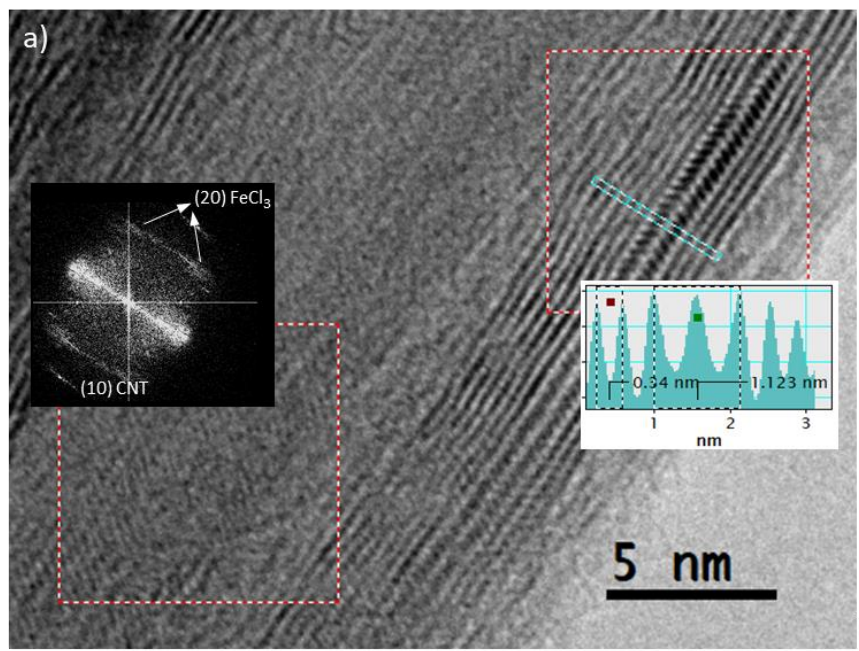
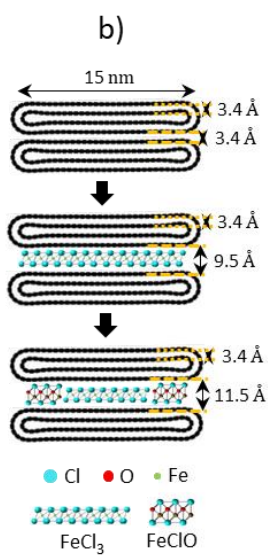
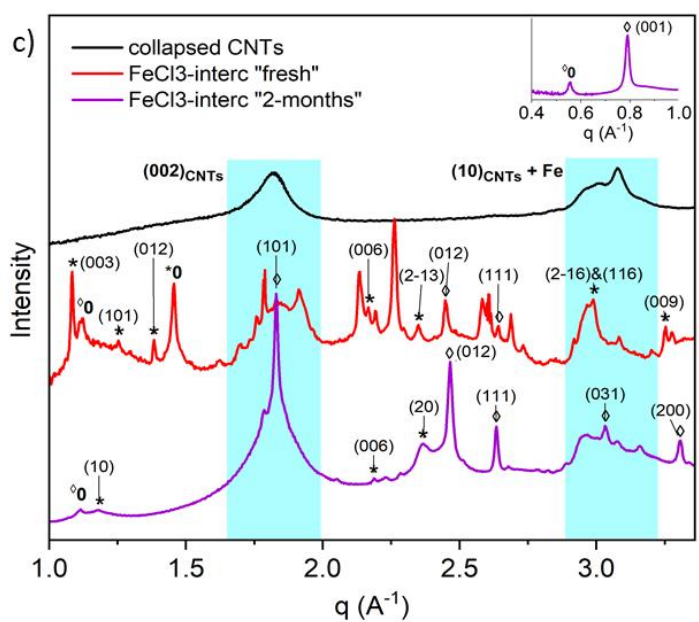

**Figure 3.** Evidence of FeCl$_3$ and FeClO in intercalated bundles. a) HRTEM micrographs show the presence of FeCl$_3$ intercalated throughout the bundles and FeClO only at bundle edges (see FFT). Spacing between graphitic planes increases to ~11.25 Å in the presence of a FeClO monolayer. b) Scheme showing structural evolution from the pristine material to the FeCl$_3$-intercalated and finally to FeCl$_3$-intercalated bundles passivated at the edges by FeClO. c) XRD patterns of the pristine collapsed CNT fibre (black), an FeCl$_3$-intercalated fibre with minimised exposure to air (red) and an intercalated samples left in ambient conditions for two months (violet). Peaks indexed with a star (*) correspond to FeCl$_3$ (R-3 space group), with a diamond (◊) to FeClO (Pmnm space group), and circles (0) to reflections of the intercalation compound. XRD data confirm that some FeCl$_3$ intercalates as multilayers and that prolonged exposure to air increases FeClO content.

Although in XRD peaks coming from interlayer FeClO reflections are clearly stronger than those from FeCl$_3$, we note that diffraction intensity is not a direct indication of the relative volume fraction of the two materials, particularly because of the atomic-thickness of most FeCl3 intercalant layers, which implies that larger FeClO structures have a more prominent contribution to XRD intensity. Indeed, under TEM, the sample exposed to air presents a large fraction of FeCl$_3$ and predominant location of FeClO at bundle edges (same situation as in **Figure 3**.a), being also many edges still intercalated by FeCl$_3$ (see Figure S4). This is sign of a remarkable stability. The large fraction of intercalated FeCl$_3$ in exposed samples is also confirmed by XPS and Raman spectroscopy, as discussed below. We also note that neither type of sample shows peaks from the typical residues found in FeCl$_3$-GICs when exposed to air for long periods (Fe$_2$O$_3$, Fe$_3$O$_4$, or FeCl$_2$·2H$_2$O) [39]. Similar stability has also been observed in FeCl$_3$ intercalated inside folded graphene layers [41], stable upon days of exposure to humid air and upon large temperatures of 620 °C in vacuum, and in high quality mesophase-pitch carbon fibers [32] and few-layers graphenes [33]. Two common features in these systems, a low defect density and high planarity to host the intercalant and shield it, are likely responsible for decreasing deintercalation rate.

### 3.2. Charge transfer in intercalated fibres

XPS measurements confirm the successful doping of collapsed CNT bundles, starting with the observation of iron and chlorine in intercalated samples both after preparation and after air exposure. A surface chemical analysis based on XPS indicates that the atomic proportion of C, O, Fe and Cl in the fresh intercalated samples are 79.9 at%, 11.4 at%, 2.0 at% and 6.7 at%, respectively. For comparison, the pristine material has corresponding atomic fractions of C, O, and Fe of 95.9 at%, 4.0 at%, and 0.1 at%, respectively. The iron signal present in the pristine collapsed CNTs comes from the catalyst particles. A detailed analysis of the XPS spectra of CNT fibres and the role of Fe catalyst and C impurities can be found in reference [42]. **Figure 4** shows XPS of pristine and intercalated samples in the C 1s, O 1s, Fe 2p and Cl 2p regions, along with peak fitting Voigt-type curves, used to study their chemical environment through the characteristic peak positions and shapes. Upon $FeCl_3$ intercalation, the C1s peak slightly broadens and its binding energy is downshifted by 0.25 eV with respect to the pristine material (**Figure 4**.a). This shift is strong indication of electron transfer from the CNTs to $FeCl_3$ in the intercalated state, consistent with observations on p-doped GICs [43,44]. It is the downshift in Fermi edge that causes reduction of the energy required for photoemission, thus leading to an apparent reduction in binding energy [43].

The XPS signals of the intercalated samples in the Fe 2p region (**Figure 4**.b) have profiles that clearly resemble that of $FeCl_3$ [45], for both the "fresh" and 2-months air exposed samples. A significant deviation in binding energy of the Cl $2p_{3/2}$ peak from 198.8 eV (in normal $FeCl_3$ crystals) [46] to 198.1 eV is encountered in the "fresh" intercalation compound (**Figure 4**.c), indicative of the charge transferred from CNTs to the $FeCl_3$ compound. This shift is typical in $FeCl_3$-GICs [44]. In the sample exposed to air for 2-months the position of the Cl $2p_{3/2}$ peak (not shown) is still lower than in isolated $FeCl_3$ crystals, indicating stable intercalation and charge transfer over long times, although with a slightly lower charge transferred than in the fresh sample. Regarding the O1s coreline (**Figure 4**.d), the 2-months air-exposed sample contains a small extra peak that can be associated with lattice oxygen in the FeClO compound [47]. Both the Cl 2p and O 1s corelines confirm charge transfer upon intercalation, preserved even after partial conversion of $FeCl_3$ to FeClO upon prolonged exposure to ambient air.

Raman spectroscopy was performed to obtain quantitative information of the charge transfer between dopant and CNTs. The Raman spectra of pristine and $FeCl_3$-intercalated collapsed CNTs samples are presented in **Figure 5** for excitation wavelengths of 532 nm and 785 nm. The Raman spectra of the pristine collapsed CNT shows significant differences with respect to conventional round DWCNTs: (i) there are no RBMs (not shown), as expected; (ii) the transverse optical (TO) and longitudinal optical (LO) modes of the G band are seemingly degenerate and form a single band; and (iii) the 2D peak is not split [48]. The absence of splitting in the 2D band also indicates that no AB Bernal stacking occurs [49].

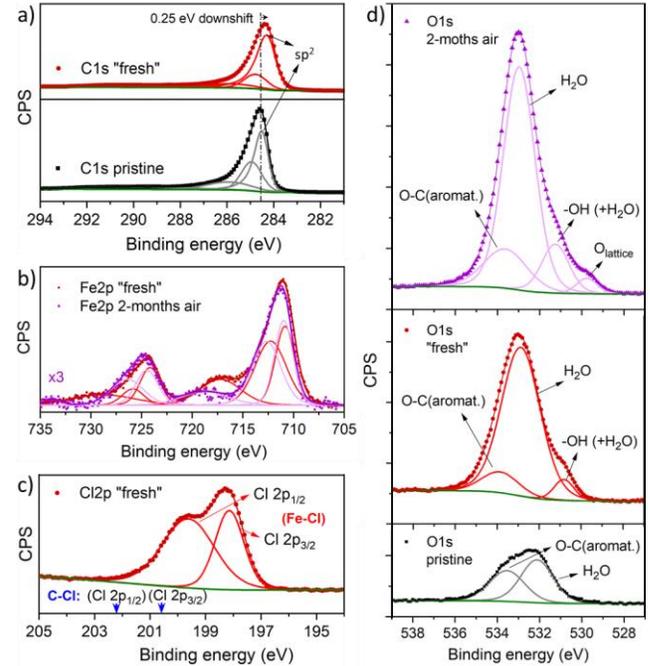

**Figure 4.** XPS spectra comparing the pristine collapsed CNTs, a $FeCl_3$-intercalated "fresh" sample (barely air exposed) and that after 2-month air exposure. a) C 1s line of the pristine and "fresh" samples; b) Fe 2p lines of the "fresh" and 2-months air exposed sample; c) Cl 2p lines of the "fresh" sample; and d) O 1s line of the three samples.

In the intercalated sample, the G band intensity is lower and both the G and 2D bands are upshifted and broader. The G band is fitted with two peaks, $G_{in}$ and $G_{out}$. We assign the two main peaks to the inner and outer layers of intercalated CNTs (see Schematic in **Figure 3**). In analogy to GICs (of stage >2), they correspond to the interior and bounding layer next to the intercalant, respectively, which have different degree of charge transfer [50–52]. The intercalated materials show an average shift in G frequency respect to the pristine material of $\Delta G_{out}=25.3\pm2.4$ cm$^{-1}$ for both laser wavelengths, and of $\Delta G_{in}=10.6\pm1.3$ cm$^{-1}$ for the 532 nm laser and $6.2\pm1.8$ cm$^{-1}$ for the 785 nm excitation. These large shifts leave no doubt of the extensive charge transfer in the system, which largely remain upon long-term exposure to ambient conditions (as shown in the histogram in Figure S5). Chlorination of the carbon walls, which could produce a G-band upshift, is discarded through the absence of the two Cl 2p lines of C-Cl bonds in XPS [53]. For comparison, in stage>2 GICs (i.e., with at least three carbon layers between intercalant layers) intercalated with $FeCl_3$, $\Delta G_{bounding}$ is around 24 cm$^{-1}$ [50], remarkably similar to the value obtained for these collapsed CNTs.

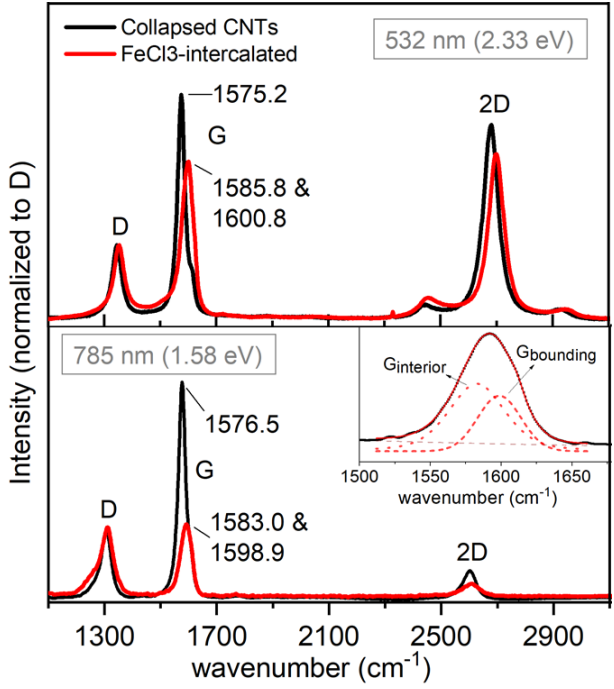

**Figure 5.** Raman spectra of the pristine and FeCl$_3$-intercalated collapsed CNT fibres taken at excitation wavelengths of: a) 532 nm (2.34 eV), and b) 785 nm (1.58 eV). All patterns are normalized to the D peak maximum intensity. Separating contributions to the Raman spectra from internal and outer layers through the use of different laser lines and multi-peak fitting of the G band enables determination of strain-like and dynamic components to the G band upshift upon intercalation.

The interest then, is in quantifying charge transferred and Fermi level shift from Raman shifts. This requires separating the two contributions to the G band shift upon intercalation [22,54]: 1) softening (hardening) of the G mode from modification of the spring force constant between carbon bonds of the sp2 lattice due to addition (removal) of electrons to (from) the antibonding (bonding) orbitals upon electron (hole) doping (referred to here as a strain contribution); and 2) a dynamic component causing upshift (downshift) of the G band, which accounts for the non-adiabatic removal (introduction) of the Kohn anomaly present at Γ at zero doping (and at band crossings at Γ in general) [55]. This dynamic component magnifies the effect of doping due to modification of the electron-phonon coupling near the Fermi level and causes renormalization of the 'bare' phonon frequency [21,56–58].

One difficulty at quantifying charge transfer and Fermi level shift from the G band upshift is compounded by the fact that the charge transferred per carbon atom ($f$) is not uniformly distributed between graphitic layers. In GICs, for example, Chacón-Torres et al. [59], calculated that around 15% of the total charge comes from the interior layers and the rest from the bounding layers directly in contact with intercalant.

In the case of DWCNTs, Chen et al. [60], found a similar value of 10% of the total charge being transferred by the inner wall. Drawing from recent work on graphene, we use existing relations between the G and 2D Raman peak positions to separate pure strain and doping contributions to the shifts [61,62], and then calculate their contributions to total charge transfer. Pure p-doping is known to produce a slope of Δ2D/ΔG of 0.55-0.75, whereas hydrostatic compression gives Δ2D/ΔG of 2.2.

**Figure 6**.a presents a plot of G vs 2D position for the pristine and intercalated samples, showing indeed a contribution with a slope similar to pure hydrostatic strain and another from p-type intercalation (strain plus dynamic effects). We are able to separate the inner and outer G-2D positions of the intercalated sample based on the fact that at high excitation energies (532 nm laser), Raman resonance is dominated by the bounding outer layers [63], rather than by the highly strained inner layers; whereas under lower excitation energies (785 nm) the contribution of the highly doped outer layers to the 2D band diminishes or even disappears due to the large shift in Fermi edge [52]. The slope dataset obtained for the inner layers of the intercalation compound is very similar to that of pure hydrostatic strain; however, we also observe a contribution from their own charge transfer. In contrast, a much larger G band upshift is obtained for the outer layers, resulting in a slope of 0.75, which is in the range of p-doped GICs, and indicative of strong dynamic effects.

From these results, we are able to calculate charge transfer between CNTs and intercalant. The G-shift components of the inner and outer CNT layers are expressed as follows:

$$\Delta G^{out} = \Delta G_\epsilon^{out} + \Delta G_{dyn}^{out}$$

$$\Delta G^{in} = \Delta G_{\epsilon,out}^{in} + (\Delta G_\epsilon^{in} + \Delta G_{dyn}^{in})$$

Where $\Delta G_\epsilon^{in/out}$ and $\Delta G_{dyn}^{in/out}$ are, respectively, the strain and dynamic contributions of charge transfer, and $\Delta G_{\epsilon,out}^{in}$ is the strain produced in the inner layers by the strained outer walls (this contribution is not present in GICs).

The intersection of the data obtained for the inner layers (at 532 nm excitation) with the strain slope in **Figure 6**.a gives $\Delta G_{\epsilon,out}^{in}$ =4.4 cm$^{-1}$. In DWCNTs, the hydrostatic strain in the inner layers caused by the strained outer layers follows [64]:

$$\Delta G_\epsilon^{out} = 1.7\, \Delta G_{\epsilon,out}^{in}$$

This gives the possibility to extract the strain contribution of doping to the G-shift associated to the outer layers; in our case, $\Delta G_\epsilon^{out}$ =7.5 cm$^{-1}$. The dynamic contribution to the G band upshift for the outer layers can then be trivially calculated. The average total G-shift of the outer layers is $\Delta G^{out}$=25.3 cm$^{-1}$, leading to a dynamic part $\Delta G_{dyn}^{out}$ of 17.8 cm$^{-1}$. (Note that this method takes into account both sources of strain induced in inner layers: by outer layer strain ($\Delta G_{\epsilon,out}^{in}$) and by doping of the inner layer; thus avoiding overestimation of charge transfer)

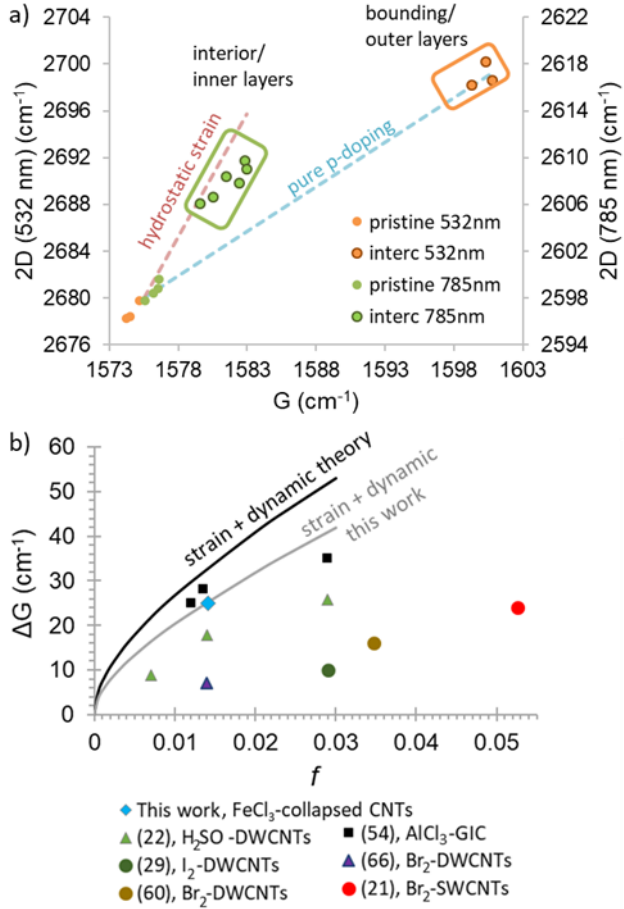

**Figure 6.** a) Raman G-2D representation. The G and 2D bands of the pristine collapsed CNTs are well fitted with a single Lorentzian. By using the 785 nm laser, we assume that the 2D transition of the highly doped outer layers does not occur (see text) and thus only the 2D peak of the inners is observed in the intercalated sample. At 532 nm excitation, however, the 2D of the outer walls are much more resonant than the highly strained inner layers. This way the G-2D was separated in inner and outer layers for the intercalation compound. In order to match the two vertical axis accounting for the different excitation wavelengths used in our experiment we using a 2D mode dispersion of 110 cm$^{-1}$/eV. b) Representation of the parametrized curves of strain (with expanded lattice) plus dynamic effects (black) obtained theoretically by Lazzeri and Mauri for graphene [65], together with experimental data from this work (and proposed parametrized curve, in grey) and from literature.

To determine charge transferred from the Raman shift components, we note that in graphitic systems the deformation of the graphene layers upon electron transfer to the intercalant behaves as a hydrostatic pressure, with G scaling in a fairly linear manner with charge transfer per carbon atom, $f$, while the dynamic part scales as the square root of $f$ [22,54,65]:

$$\Delta G = \Delta G_\epsilon + \Delta G_{dyn} = a\,f + b\,\sqrt{f}$$

Lazzeri and Mauri obtained the theoretical $\Delta G$-$f$ tendencies for graphene in two cases [65]: 1) applying adiabatic DFT plus expanded lattice effects (thus, the strain contribution of charge transfer can be extracted); and 2) applying time-dependent perturbation theory (TDPT) plus expanded lattice effects, so the sum of strain and dynamic contributions is obtained. By digitalizing the graph therein obtained, we estimate the factors $a$ and $b$ to be 530 and 214, respectively. Experiments show quite good agreement for graphite intercalated with $AlCl_4^-$ [54].

By assuming that the strain factor, $a$, in collapsed CNTs is equal to that of graphite, we obtain an $f$ ($= \Delta G_\epsilon^{out}/a$) of 0.014 electrons per carbon atom, equivalent to a surface electron concentration of $\sim 5.3 \cdot 10^{13}$ cm$^{-2}$ for the outer layer or to a donated electron per 70 carbon atoms. Factor $b$ can then be determined from knowledge of $f$ and the components of $\Delta G$, yielding the complete relation between Raman shift and charge transferred in the p-type intercalated collapsed CNTs as:

$$\Delta G = 530\,|f| + 150\,\sqrt{|f|}$$

In **Figure 6**.b, we plot experimental G-shifts for different intercalation compounds of graphitic materials and acceptor compounds, as well as charge transfer calculated using alternative simpler methods. The results show that the method proposed here for CNTs gives a correspondence between G-shift and charge transfer close to GICs, whereas other report calculate unusually high charge transferred for a given G-shift.

Finally, we use expressions derived from density functional theory to relate charge transferred by the outer layers to their shift in Fermi edge [65,66]:

$$\Delta E_{F,out} = -6.04\,\sqrt{|f|} = -0.72\ \text{eV}$$

Note that the large Fermi level shift obtained supports the strategy used to discriminate inner and outer layers through use of different laser excitation energies[1].

---

[1] This value validates the approach employed in Raman that using the 785 nm laser ($E_L$=1.58 eV), the 2D peak obtained entirely corresponds to the inner layers. For the transition to occur (not being forbidden by Pauli blocking), the shift in Fermi edge must fulfil that $|\Delta E_F|<[E_L-\hbar\omega_{2D}]/2$ [33,59]. In this case, the obtained Fermi edge shift for the highly doped outer wall is larger than the right side of the expression at 785 nm excitation (that gives 0.63 eV).

Finally, in order to obtain a first indication of the effect of charge transfer on the bulk properties of the CNT fibre fabrics and single filaments, we performed atmospheric 4-probe electrical resistance measurements on a sample before and after intercalation. The electrical conductivity increased by a factor of 6.6 after intercalation with $FeCl_3$.

## 4. CONCLUSIONS

This work presents the first demonstration of vapour-phase intercalation in bundles of collapsed CNTs, using $FeCl_3$ as intercalant. Direct imaging of intercalated areas through TEM shows that $FeCl_3$ forms ordered long length domains (from tens to hundreds of nanometres), mostly as monolayers extended between the collapsed CNT ribbon-like crystals. Exposure to ambient air leads to formation of FeClO at some bundle edges, where it acts as a stabiliser by preventing deintercalation or further hydrolysis of the inner $FeCl_3$ layer. XPS measurements confirm electron transfer from the CNTs to $FeCl_3$ upon intercalation, which remains even after months of exposure to ambient conditions. The Raman spectra of intercalated samples show a large upshift of the G band by up to 25 $cm^{-1}$, corroborating the large charge transfer obtained with the $FeCl_3$ metal halide as intercalant. Using different excitation energies enabled separation of the spectra for inner and outer layers of the CNTs, assigned to the interior and bounding layer of the intercalated system, which allowed us to correctly extract the value of the G-shift corresponding to the strain contribution, from where the charge transfer per carbon atom, $f$, is obtained. The procedure enabled accurate determination of charge transfer and Fermi level shift, by taking into account the theoretical (and experimentally supported) fact that $\Delta G$ follows a linear tendency with $f$ for the strain contribution and a square root tendency for the dynamic contribution. For the outer CNT layers in contact with intercalant, 0.014 electrons are transferred per C atom, equivalent to a surface electron concentration of ~$5.3 \cdot 10^{13}$ $cm^{-2}$. The corresponding shift in Fermi level is –0.72 eV. A comparison with reported data for G-shift for different intercalated carbons and the charge transferred calculated by different methods shows that the approximation proposed here is in agreement with GIC work, whereas previous reports on CNT appear to highly overestimate charge transferred. Alternative techniques to directly quantify charge transferred are needed to clarify this discrepancy. Finally, doping of the collapsed CNTs leads to a 6-fold increase in longitudinal conductivity.

Future work is directed at comparing the kinetics and stability of intercalated collapsed CNTs in relation to round CNTs and GICs, and in studying in detail the transport properties of these intercalated yarns.

## ACKNOWLEDGEMENTS

The authors are grateful for generous financial support provided by the European Union Seventh Framework Program under grant agreement 678565 (ERC-STEM), by the MINECO (RyC-2014-15115), by the Air Force Office of Scientific Research of the US (NANOYARN FA9550-18-1-7016), and by "Comunidad de Madrid" FotoArt-CM project (S2018/NMT-4367). M.V. acknowledges the Madrid Regional Government (program "Atracción de Talento Investigador", 2017-T2/IND-5568) for financial support.

## APPENDIX A. SUPPLEMENTARY DATA

Supplementary data to this article can be found online at: https://

# Supplementary information - Macroscopic yarns of FeCl$_3$-intercalated collapsed carbon nanotubes with high doping and stability

Cristina Madrona,[1,2] María Vila,[1] Freddy E. Oropeza,[3] Víctor A. de la Peña O'Shea,[3] Juan J. Vilatela[1,*]

1 IMDEA Materials Institute, Eric Kandel 2, 28906 Getafe, Madrid, Spain

2 Facultad de Ciencias, Universidad Autónoma de Madrid, Francisco Tomás y Valiente, 7, 28049 Madrid, Spain

3 IMDEA Energy Institute, Avda. Ramón de la Sagra 3, 28935 Móstoles, Madrid, Spain

*KEYWORDS Collapsed carbon nanotubes, vapour-phase, intercalation compound, acceptor dopant, fibre, charge transfer.*

## 1. TEM MICROGRAPHS OF PRISTINE VS INTERCALATED COLLAPSED CARBON NANOTUBES

HRTEM images of the carbon host, consisting on collapsed CNTs forming bundles, are presented in **Figure S1**.a and **Figure S1**.b. The interlayer distances are quite homogeneous, as observed in the intensity line profiles taken at specific regions marked in the images. By other side, in **Figure S2**, which corresponds to part of the HRTEM micrograph shown in.a, the appearance of FeCl$_3$ monolayers is tracked by the exposed line intensity profile, which shows distances between carbon layers of around 9.2 Å whenever a monolayer of FeCl$_3$ is intercalated.

## 2. SAED OF PRISTINE VS INTERCALATED BUNDLES

Selected area electron diffraction (SAED) measurement in oriented bundles of the pristine host material is shown in **Figure S1**.c, with part of its selected area shown in **Figure S1**.d. Reflexions coming from this oriented graphitic structure are indexed. By other side, a SAED pattern in oriented bundles of the intercalated material is taken in a zone where both monolayers and multilayers of FeCl$_3$ are intercalated. Some reflexions of the graphitic planes (G) and of ferric chloride (F) are indexed. The reflection at q=1.48 Å$^{-1}$ (d=4.24 Å), appearing also in wide-angle x-ray scattering (WAXS) measurements, is aligned with diffraction spots coming from the (002) of the CNTs and the (003) of FeCl$_3$ (see SAED in **Figure S3**.a, corresponding to the selected area appearing partially in **Figure S3**.b). We assign this peak to the distance between the outermost FeCl$_3$ layer (when multilayers are intercalated) and its adjacent carbon layer, which is supported by direct measurements of this distance in HRTEM micrographs (see **Figure S3**.c). This distance of ~4.3 Å is interestingly much lower than the sum of half the interlayer distances of FeCl$_3$ (5.65 Å in the intercalated state and 5.8 Å in normal FeCl$_3$ crystals) and of the graphitic planes (3.45 Å). This decreased distance may result from a high interaction between FeCl$_3$ and carbon layers in the intercalated state.

## 3. TEM MICROGRAPHS OF SAMPLES AFTER EXTENDED EXPOSURE TO AMBIENT AIR

HRTEM images of the FeCl3-intercalated collapsed CNT sample exposed to air for two months are presented in **Figure S4**. It is important to note that many regions with FeCl3 are still found at the edges in this 2-months air exposed sample, evidencing the remarkable stability of the intercalation compound.

## 4. RAMAN G-BAND EVOLUTION UPON EXTENDED EXPOSURE TO AIR

In order to show stability of the intercalated samples, the evolution of the Raman G band of the inner and outer walls ($G_{in}$ and $G_{out}$) is presented as a histogram in **Figure S5** for different periods of exposure to ambient air over months. A gradual decrease in G-shift with respect to the pristine material is observed, indicating a loss of charge transfer over time, but even after 8 months the material remains highly doped.

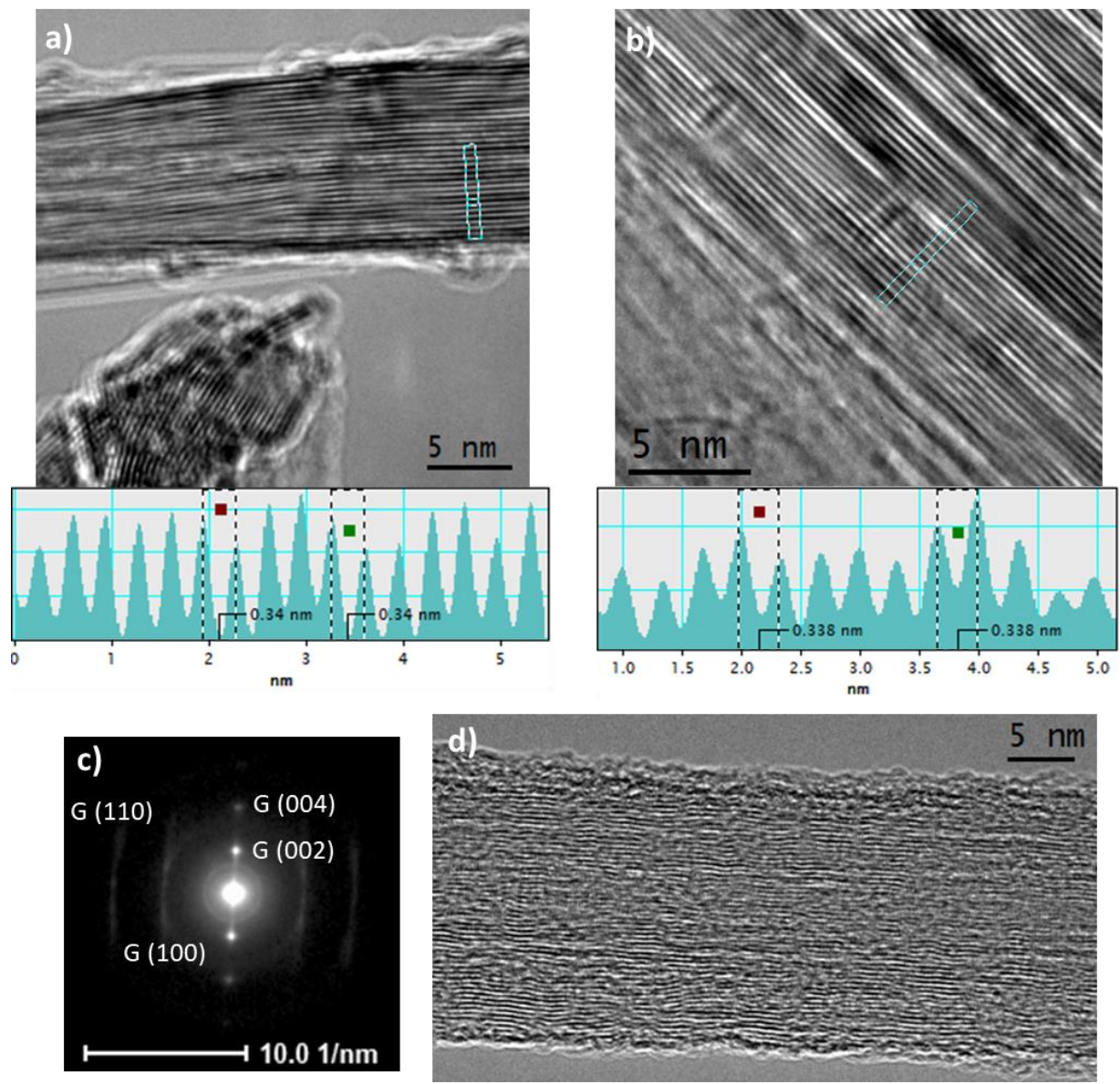

**Figure S1**. a)-b) HRTEM micrographs of the pristine (unintercalated) collapsed CNTs showing the interlayer distances measured directly through a line intensity profile. In a) the collapsed structure of the CNTs can be observed in the bottom part. c) Selected area electron diffraction (SAED) pattern obtained in an oriented bundle of the pristine collapsed CNTs; reflexions of the graphitic planes (G) are indexed. d) HRTEM micrograph showing part of the selected area where the SAED pattern in c) was taken.

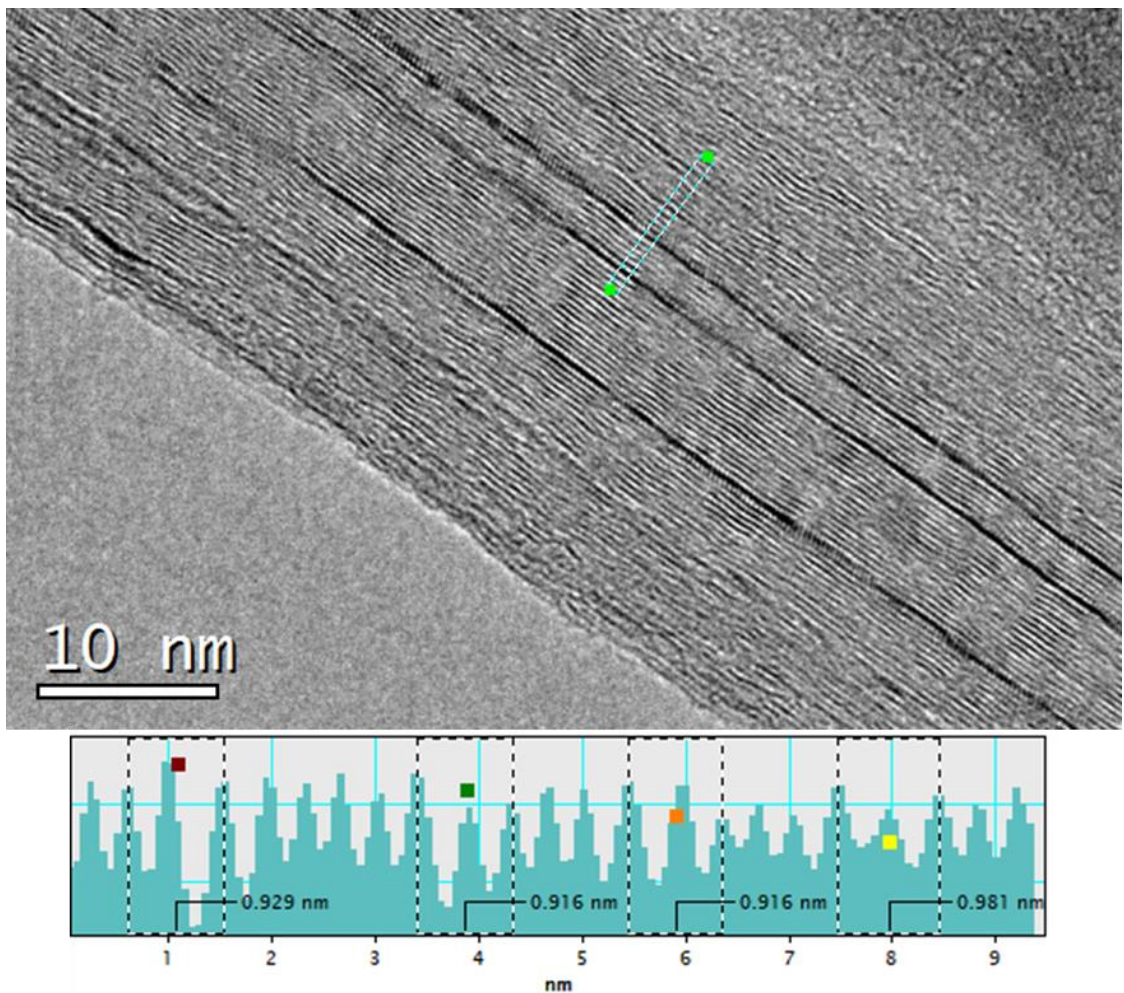

**Figure S2.** HRTEM micrograph corresponding to an amplification of Figure 2.a presented in the main article. The measured distances show the appearance of stage 4 $FeCl_3$-collapsed CNTs in part of this region (i.e., one intercalant layer every four carbon layers).

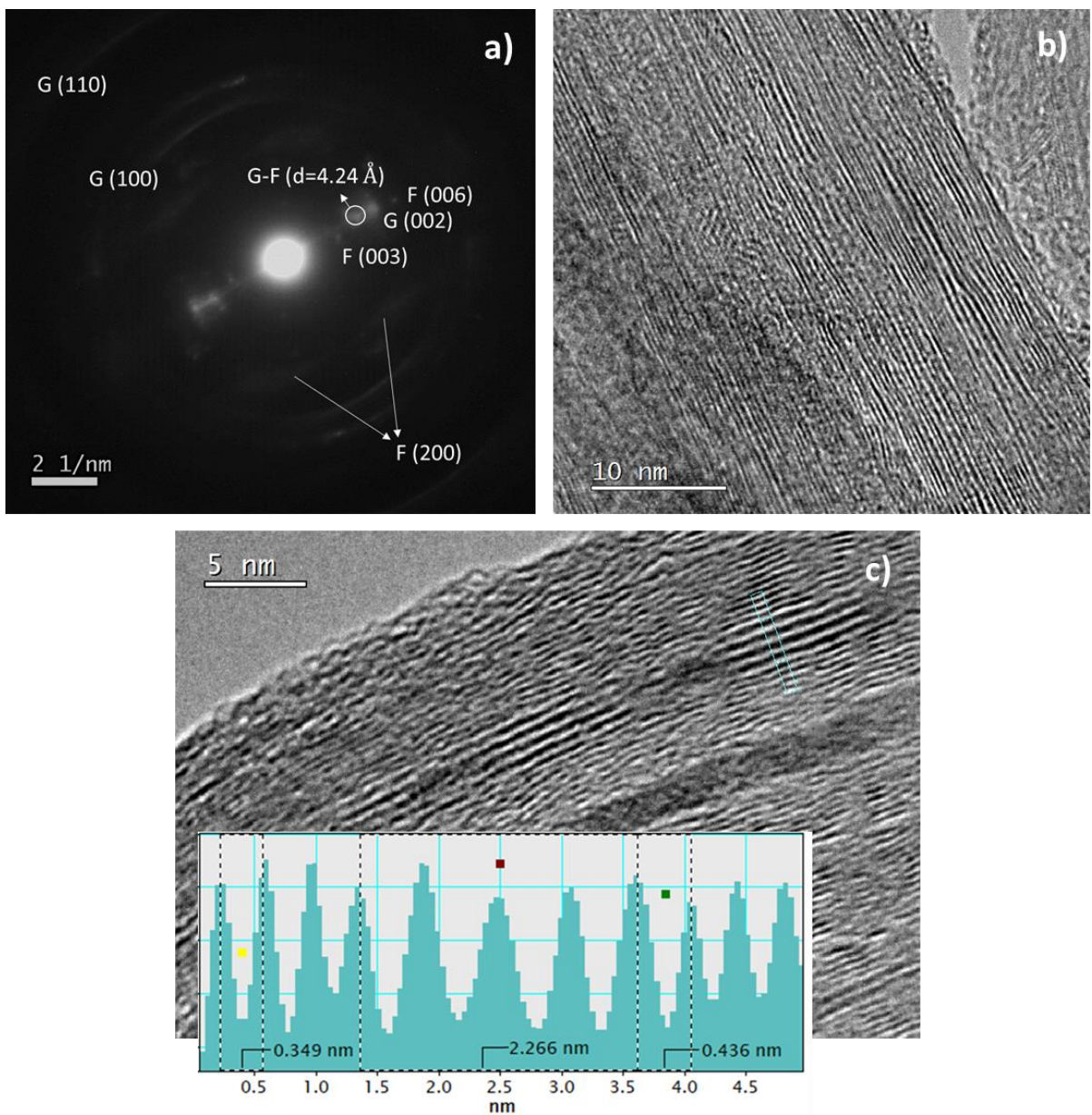

**Figure S3.** a) Selected area electron diffraction (SAED) pattern obtained in oriented bundles, in a zone where both monolayers and multi-layers of $FeCl_3$ are intercalated. Some reflexions of the graphitic planes (G) and of the ferric chloride (F) are indexed. We assign the reflection at d=4.24 Å (aligned with the (002) interplanar reflection of the CNTs) to the distance between the outermost $FeCl_3$ in multilayers and its adjacent carbon layer (thus indexed as F-C). b) HRTEM micrograph showing part of the selected area where the SAED pattern in a) was taken. c) Interplanar distances measured directly in the exposed HRTEM micrograph; distance between graphitic planes (yellow, ~3.49 Å), between five $FeCl_3$ intercalated planes (red, giving an average interplanar distance of ~5.65 Å), and between the outermost layer of $FeCl_3$ and its adjacent carbon layer (green, ~4.36 Å).

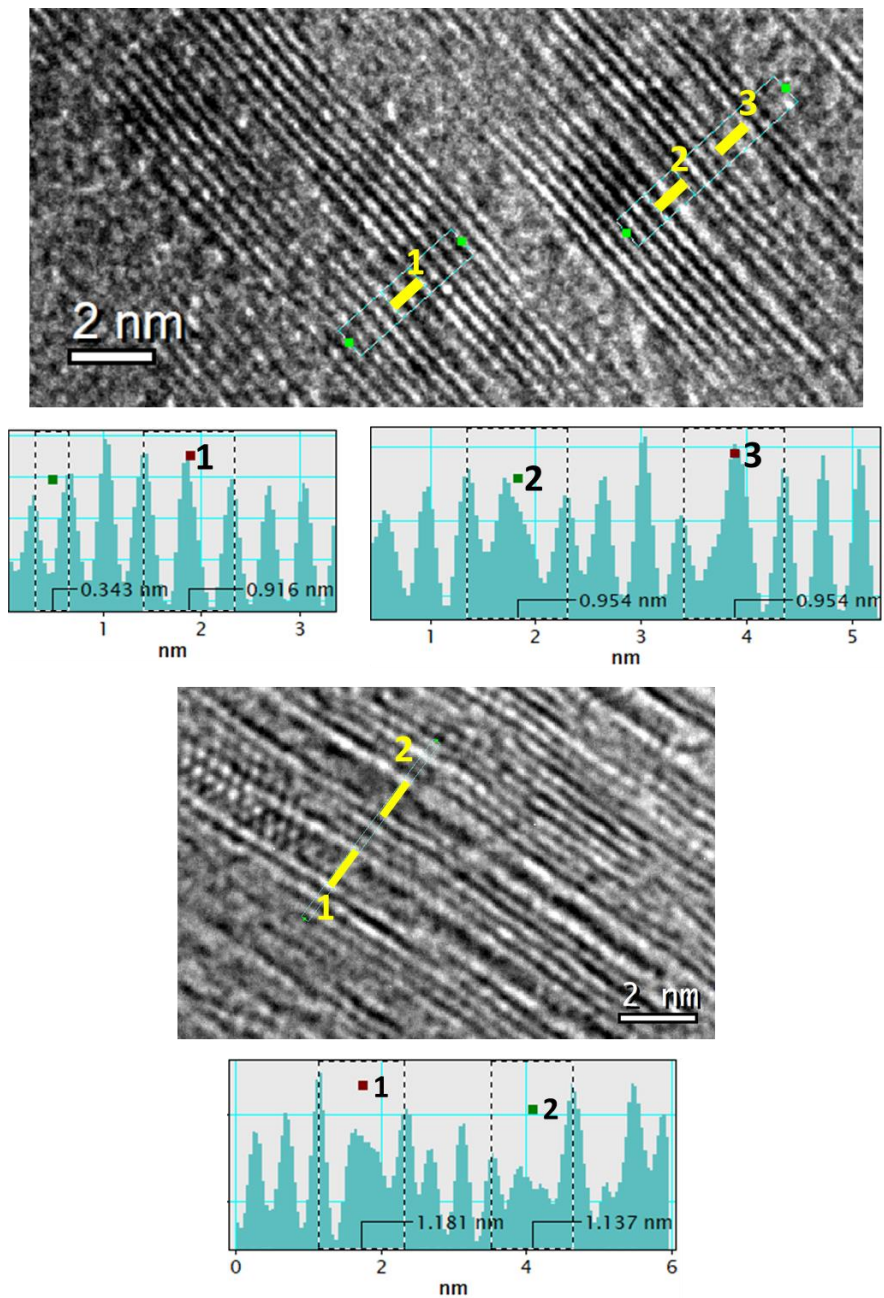

**Figure S4.** High resolution TEM micrographs of the sample exposed to air for two months, showing zones with FeCl$_3$ at the edges (upper image) and others with FeClO at the edges (bottom image). Notice that many regions with FeCl$_3$ are still found at the edges, evidencing the remarkable stability of the FeCl$_3$-collapsed CNT compound.

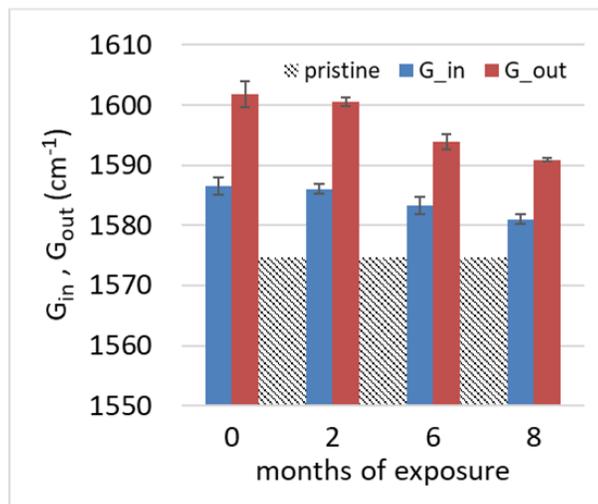

**Figure S5.** Evolution of the Raman G band of the inner and outer walls ($G_{in}$ and $G_{out}$) through months of exposure to ambient air. The decrease in G-shift with respect to the pristine material upon time of exposure indicates a loss of charge transfer over time. Measurements are obtained with a 532 nm excitation laser.